\documentclass[12pt]{article}
  \usepackage{a4}
  \newcommand\ignore[1]{}
 \usepackage{epsfig,graphicx}
\begin{document}
 
 \hfill To be published in Proceedings of ECRYS-99,

\hfill Journ. de Physique, Coll., December 1999.

\vskip .6in

  {\Large{\bf{Topological coupling of dislocations and magnetization vorticity in Spin Density Waves.}}}
 \vskip .1in
 {\centerline{S. Brazovski$^{1,2}$,  N. Kirova$^{1}$.}
  \vskip .1in

{\it{ $^1$Laboratoire de Physique Th\'eorique et des Mod\`eles Statistiques, CNRS,}

\it{$^{\ }$B\^at.100, Universit\'e Paris-Sud, 91406 Orsay cedex, France.}

\it{$^2$L.D. Landau Institute,  Moscow, Russia. }}

 \parskip 6pt

  \vskip .2in

  \begin{narrow}
  {\small
  {\bf Abstract.}
The rich order parameter of Spin Density Waves allows for an
unusual object of a complex topological nature:   a half-integer dislocation combined with a semi-vortex of the staggered magnetization. It becomes energetically preferable to ordinary dislocation  due to enhanced
Coulomb interactions in the semiconducting regime. Generation of these objects changes the narrow band noise frequency.}
 \end{narrow}

\vskip .2in
{\bf 1.  Introduction.}
\vskip .1in
Topological  defects in Electronic Crystals - solitons, phase slips (PS) and dislocation lines/loops (DLs) are  ultimately necessary for the current
conversion and depinning processes. 
 Microscopically  in Charge and Spin Density Waves (CDW, SDW, DW) the PS starts as a self-trapping of electrons into solitons with their subsequent aggregation (see [1] and references therein). Macroscopically the  PS develops as the edge  DL proliferating/expanding  across the sample [2]. At low temperatures $T$, the energetics of DL in DW  are determined by the Coulomb forces [3] and by screening facilities of free carriers. The CDW/SDW are characterized  by scalar/vector order parameters:  $\eta_{cdw}\sim\cos[Qx+\varphi]$,  $\vec\eta_{sdw}\sim\vec m \cos[Qx+\varphi]$ where $\vec m$ is the unit vector of the staggered magnetization.   Here we will show that  SDW allow for unusual $\pi$ PSs forbidden in CDW where only $2\pi$ PSs are allowed (see also [4]). Namely in SDW conventional dislocations loose their priority in favor of special topological objects: a half-integer dislocation combined with a semi-vortex of a staggered magnetization vector. The magnetic anisotropy confines   half-integer DLs in pairs connected by a magnetic domain wall. As a possible manifestation, the $\pi$- PSs reduce twice (down to its CDW value  $\Omega /j=\pi $) the universal ratio $\Omega /j$ of the fundamental
frequency $\Omega $  to the mean $dc$ sliding current $j$. The splitting of the normal $2\pi$-dislocation to the $\pi$ ones in energetically favorable due to  Coulomb interactions.

\vskip .2in

  {\bf 2. A single dislocation in semiconducting density wave.}
\vskip .1in

Consider a D-loop of a radius  $R$  embracing a number $N=\pi R^2/s$ chains 
($s=a_{\bot }^2$ is the area per chain) or a D-line at a distance $R=a_\bot N$ from its counterpart or from the the surface. 
 The primary energy scale $E_0$ of DL is the measure of the interchain coupling. For CDW it is  
 usually $E_0\sim T_c $  while for typical  SDW with  a strong electronic overlap  $t_\bot$,   
$E_0\sim t_\bot$ which can be larger than $T_c$. The energy of a pure magnetic vortex loop is not affected by the Coulomb forces, so its only scale is  $E_0$. But for DLs the compressibility is involved with respect to the phase deformations which are charged and hence greatly affected by Coulomb forces. Namely, the compressibility  hardens with $R$  as $R^2/r_0^2$ for $R$ beyond the screening length of the parent metal $r_0\sim 1\AA$, until it saturates at the actual screening length $r_{scr}=r_0/\sqrt{\rho _n}$, where $\rho_n$ is the normal density due to carriers activated through the DW gap $2\Delta$.
Finally for the D-loop energy ${\cal H}(N)$ one find [3]:

$$
{\cal H}(N)\sim \sqrt N \ln N E_0 \ \ \ \ \ \ \qquad \qquad  N\sim 1 \  {\rm for \ DLs; \ all}\  N \ {\rm for \ vortex \ loops}
$$
$$
{\cal H}(N) \sim \  N E_0 a_\bot /r_0 \ \ \ \ \ \ \qquad \qquad  R<r_{scr} \qquad (N=\pi R^2/s) \qquad\qquad\quad
$$
$$
{\cal H}(N) \sim  \sqrt N \ln N  E_0 r_{scr}/r_0 \quad \qquad R>r_{scr} \qquad\qquad\qquad\qquad \qquad\qquad\quad
$$

We see that within $r_{scr}$ there is no  perimentrical law but rather the area one for the dislocation energy. At large distances the standard $ \sqrt N$ - law  is restored but enhanced as $\sim \rho_n^{-1/2}$. 
  The equilibrium with respect to aggregation of electrons to the  DL is controled by the chemical potential 
$\mu _D=\partial {\cal H} (N)/\partial N$.
 Resulting  $\mu_D$ of a single D-loop   is drown schematically at Fig. 1. Here the dashed line corresponds to the   magnetic vortex loop. The inner region of the solid line describes   the D-loop at shortest distances where the Coulomb forces are not important yet.   

\begin{figure}[thb]
\includegraphics{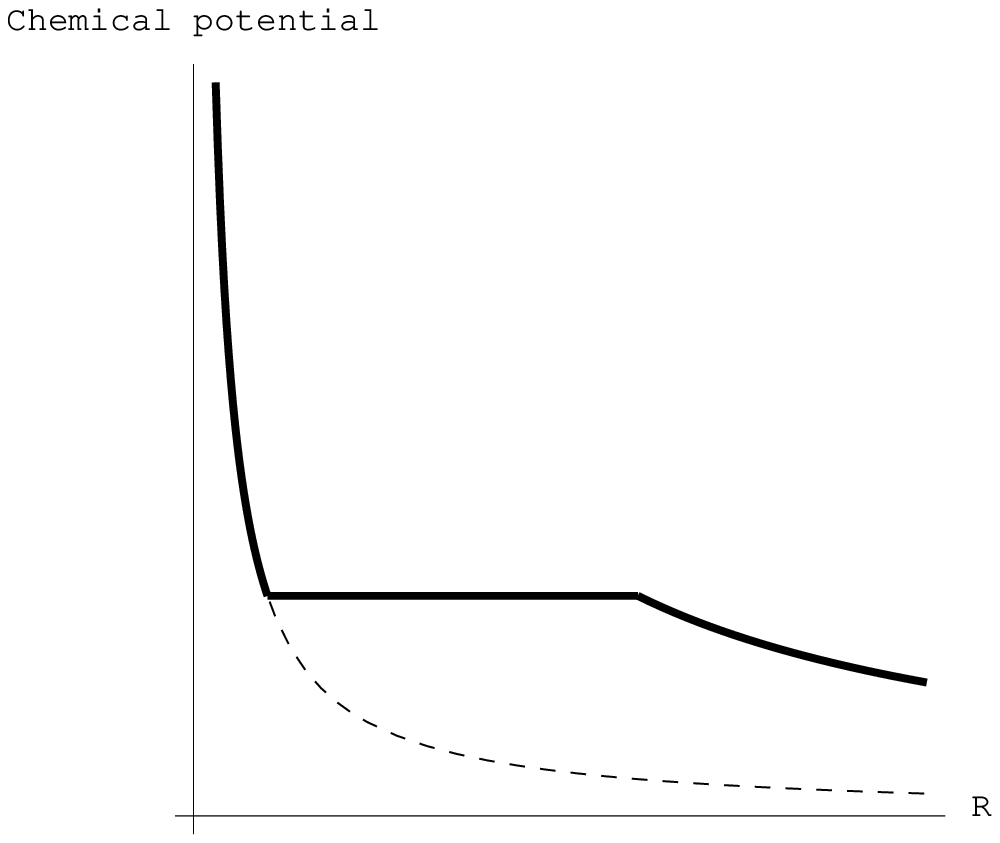}
\epsfxsize 1.5 \hsize
\caption{The chemical potential $\mu_D=\partial W/\partial N$ of the D-loop. Dashed line corresponds to the model without Coulomb interactions and to the vortex loop.}
\end{figure}
\vskip .2in
{\bf 3. Combined topological defects in SDW.}
\vskip .1in

In SDW the Coulomb enhancement  of the dislocation energy plays
a principal role  to bring to life a special combined topological
object. This is the half-integer dislocation accompanied by $180^o$ rotation
${\cal O}_\pi$ of the staggered magnetization $\vec m$. 
Indeed, the SDW order parameter  allows for the following three types of self-mapping
$\vec \eta \rightarrow \vec \eta$. (The mapping is a general requiement of topological connectivity which selects the allowed configurations [5,6].) \\
 i. normal dislocation: 
$\ \ \varphi \rightarrow \varphi + 2\pi,\ \vec m \rightarrow \vec m$;\\
 ii. normal $\vec m$ -- vortex: $\vec m \rightarrow {\cal O}_{2\pi}\vec m, \ \ \
\varphi \rightarrow\varphi$;\\
 iii. combined object : $\ \ \varphi \rightarrow \varphi + \pi, \ \ \vec m
\rightarrow {\cal O}_{\pi} \vec m = - \vec m $.\\
A necessity of semi-vortices in conventional antiferromagnets in presence of  frozen-in host lattice dislocations has been realized already in [7].
In SDW the semi-vortices  become  the objects of the lowest energy created in the course of PS process. 
Indeed, not far below $T_{c}$ at $\rho _n\sim 1$ the elastic moduli related to the phase displacements and to magnetization rotations are of the same order. Hence all three objects have similar energies. With lowering $T$ the energy  of the object ii. is not affected because charges are not perturbed so that Coulomb forces are not involved. For objects i. and iii. the major energy $\sim
\rho _n^{-1/2}$ is associated to distortions of $\varphi $  so that the
energy of $\vec{m}$ rotation in case iii. may be neglected.  To compare the main contributions to the energies of objects i. and iii. we remind that at  given $R$ the DL energy depends on its winding number $W$ as $\sim W^2$, where $W=1/2$ for $\pi$-DL and  and $W=1$ for $2\pi$-DL. We must compare their energies at the given number of accumulated electrons $2N$. For shortness consider only the largest (screened) sizes of DLs.
For the D-loop and the D-line we have correspondingly

$$N\sim WR^2, \qquad
\mu _D\sim \partial (W^2R\ln R)/\partial (WR^2)\sim W^{3/2}/N^{1/2}$$

$$N\sim WR, \qquad
\mu _D\sim \partial (W^2\ln R)/\partial (WR) \quad \sim \ W^2/N \qquad
$$

 In both cases the lowest energy per electron is
given by an object with smallest $W=1/2$ i.e. by the combined one. Thus  in SDW the normal dislocation must split into two objects of the combined topology. Apparently they will have the same sign of the displacive half-integer winding numbers and opposite signs of the half-integer spin rotation numbers. 
In Fig.2 we present the vector field of the local  SDW magnetisation $\vec \eta$ for such a hymer.  
\begin{figure}[thb]
\includegraphics{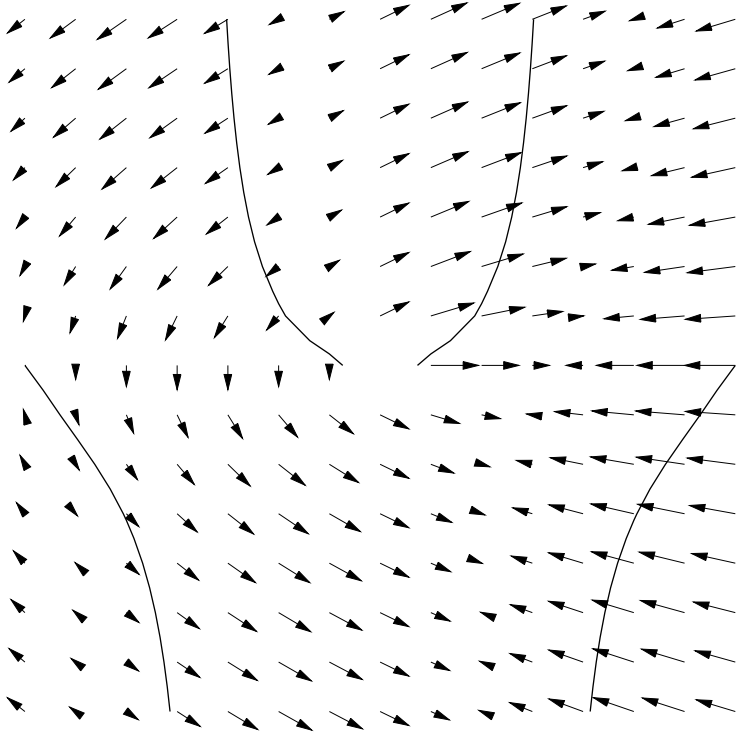}
\epsfxsize \hsize
\caption{The vector-field $\vec\eta$ for half-DL combined with semi-vortex.  Solid lines indicate  constant phases around the half-DL.}
\end{figure}
 The chain axis is horizontal. Due to the presence of the half-integer DL, the number of sites changes between the upper and the lower rows from 2.5 periods (6+6+3) to 2 periods (7+5).

In the presence of  spin anisotropy for all three orientations, the free rotation of spins is prohibited at large distances from the DL. Then  the two objects will be bound by a string which is the Neel domain wall. It can be shown to stretch along the interchain direction. Usually the spin anisotropy is noticeable only for one orientation and  characterized by the spin-flop field $H_{s-f}\sim 1T$. It would originate the string of the length $\sim 0.1 \mu m$. At higher magnetic fields
only a small in-plane anisotropy is left so that the string length may reach the  sample width.
 
We conclude  that the sliding SDW should generate "hymers": the combined topological objects where the spin rotations are coupled to the DW displacements.
The "hymers" are stable lowering the DL Coulomb energy. This combination effectively reduces the SDW period 
allowing e.g. for the twice decrease in the NBN frequency, which is an important disputable question [8,9]. 
 The interest in such unusual topological objects may go far beyond the NBN generation or the current conversion problem in SDWs.
\vskip .2in
{\bf References}

\medskip
\begin{tabular}{rl}
1.&S. Brazovskii in  {\em Proceedings of the NATO Summer School,}
 {\em Les Houches 95},\\
&C. Schlenker and 
 M. Greenblatt eds., World Sci. Publ., 1996.\\
2.&N.P.~Ong and K.~Maki, Phys. Rev. B {\bf 32}, 6582 (1985).\\
3. &S. Brazovskii and S. Matveenko,  J. de
Phys. I., {\bf 1}, 269 \& 1173 (1991)\\
4.&N. Kirova and S. Brazovsskii, Synth. Met., {\bf 103}, 1831 (1999).\\  
5.&N.D. Mermin, Rev. Mod.Phys., {\bf 51}, 591 (1979).\\
6.&V.P.  Mineev, Sov. Sc.Rev., Sec.A Phys. Rev., {\bf 2}, 173 (1980).\\
7. &I.E. Dzyaloshinskii, JETP Lett., {\bf 25}, 98 (1977).\\
8.&E. Barthel {\it et al},  J. de
Phys. I., {\bf 1}, 1501 (1991).\\
9.&W.G. Clark {\it et al}, Synth. Met., {\bf 86}, 1941 (1997).\\
\end{tabular}

\end{document}